\documentclass[prl,twocolumn,letterpaper, superscriptaddress,longbibliography]{revtex4-1}
\usepackage{graphicx}
\usepackage{dcolumn}
\usepackage{bm}
\usepackage{color}
\usepackage{tabularx}
\usepackage{array}
\usepackage[toc,page]{appendix} 

\usepackage[utf8]{inputenc}
\usepackage[T1]{fontenc}
\usepackage{color,soul}
\usepackage{tabularx}
\usepackage{amsmath}

\begin{document}
\title{An Adaptive Coherent Interferometric Oscillator Based on an Optoelectronic Magnonic Parametric Oscillator}

\author{Shihao Zhou}
\affiliation{Department of Physics and Astronomy, University of North Carolina at Chapel Hill, Chapel Hill, NC 27599, USA}

\author{Junming Wu}
\affiliation{Department of Physics and Astronomy, University of North Carolina at Chapel Hill, Chapel Hill, NC 27599, USA}

\author{Jiazhen Li}
\affiliation{Department of Electrical and Computer Engineering, North Carolina State University, Raleigh, NC 27695, USA}
\affiliation{Organic and Carbon Electronics Laboratories (ORaCEL), North Carolina State University, Raleigh, NC 27695, USA}

\author{Qing Gu}
\affiliation{Department of Electrical and Computer Engineering, North Carolina State University, Raleigh, NC 27695, USA}
\affiliation{Department of Physics, North Carolina State University, Raleigh, NC 27695, USA}
\affiliation{Organic and Carbon Electronics Laboratories (ORaCEL), North Carolina State University, Raleigh, NC 27695, USA}

\author{Wei Zhang}
\email{Email: zhwei@unc.edu}
\affiliation{Department of Physics and Astronomy, University of North Carolina at Chapel Hill, Chapel Hill, NC 27599, USA}

\date{\today}

\begin{abstract}

We study a Mach--Zehnder interferometer (MZI)-based optoelectronic magnonic parametric oscillator (OEMPO) incorporating a YIG-loaded magnonic branch and a tunable phase-shifter branch, enabling systematic investigation of adaptive interferometric oscillator dynamics under distributed phase perturbations. Through analysis of nondegenerate OEPO mode pairs and frequency-pulling behavior, the loop free spectral range (FSR) and effective delay time were quantitatively extracted. Despite the nominally frequency-pinned parametric operation, weak frequency pulling and OEPO mode softening were observed, revealing an additional adaptive interferometric degree of freedom introduced by the MZI architecture. By comparing local and global sampling configurations, we demonstrate that the YIG branch behaves predominantly as a local dispersive resonant subsystem governed by the complex magnonic susceptibility, whereas the phase-shifter branch primarily controls the global interferometric redistribution geometry. Nevertheless, coherent recombination and adaptive regeneration within the loop produce finite cross-coupling between the two branches, resulting in partially synchronized interferometric dynamics and branch-dependent adaptive redistribution. Quantitative complex-Lorentzian analysis further reveals substantial phase-to-amplitude conversion and distinct differences between the OEO and OEPO regimes: the phase-pinned OEPO favors strongly dispersive local YIG response, while the frequency-adaptive OEO exhibits more mixed absorptive--dispersive behavior due to spectral relaxation through frequency pulling. Broadly, the present platform establishes a versatile framework for exploring adaptive nonlinear interferometric physics, coherent phase redistribution, and branch-dependent synchronization phenomena in hybrid magnonic-photonic oscillator systems. 

\end{abstract}

\maketitle

\section{Introduction}

Incorporating magnonic systems into distributed coherent architectures \cite{wang2024nanoscale,lambert2025coherent,han2021microwave} such as optoelectronic oscillators (OEOs) \cite{xiong2020experimental,xiong2024magnon,wu2025coupling} has recently emerged as an attractive direction for microwave photonics, sensing, and communication technologies spanning classical and quantum regimes \cite{awschalom2021quantum,chumak2022advances,flebus20242024,barry2023ferrimagnetic,mi2026ultrasensitive,crescini2021phase}. By combining the ultra-low-loss coherent properties of optoelectronic feedback loop \cite{chembo2019optoelectronic} with the strong tunability and high sensitivity of magnonic excitations \cite{zhang2026perspective}, such hybrid systems offer promising opportunities for high-coherence microwave generation, interferometric sensing, nonlinear signal processing, and microwave-to-photonic transduction \cite{li2021tutorial,hao2023perspectives,caleffi2025quantum,zhang2026perspective}. In particular, a derivative architecture of OEO, namely, the optoelectronic parametric oscillator (OEPO) \cite{hao2020optoelectronic}, introduces additional advantages including parametric frequency pinning, coherent phase locking, nondegenerate mode-pair formation, and enhanced nonlinear gain-phase dynamics \cite{zhang2025gain,cen2022large,zhang2023hybrid,qu2025pump,hatzon2025sharp,kondrashov2022self}.  

\begin{figure*}[htb]
 \centering
 \includegraphics[width=6.3 in]{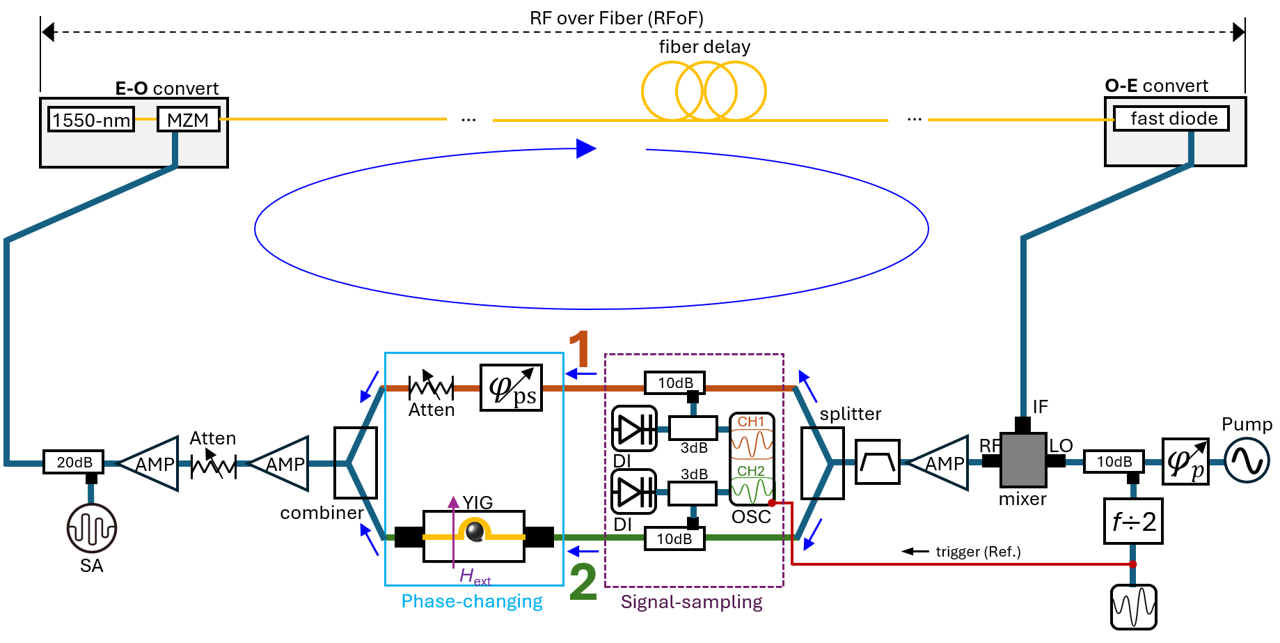}
 \caption{Schematic illustration of the setup. The top shows the photonic sector consisting of E-O / O-E converters and a fiber delay. The bottom is the microwave sector consisting of: (from right to left) the mixer RF port, amplifier (Amp), band-pass filter (1730-2270 MHz), splitter, MZI (1 $\&$ 2), combiner, Amp, attenuator (Atten), Amp, 20-dB coupler, spectrum analyzer (SA). Inside the MZI: branch 1, the reference branch, consists of a microwave phase shifter (PS) with tunable phase ($\varphi_\mathrm{ps}$) and an Atten; branch 2, the magnonic branch, consists of a YIG sphere placed atop a microwave stripline, which can induce a resonant, dispersive phase shift upon YIG resonance ($\varphi_\mathrm{YIG}$). For both branches, a 10-dB coupler is used for sampling the signal, which is further split (at the 3-dB divider) to a microwave diode (DI) for measuring the power, and to the oscilloscope (OSC) for monitoring the real-time oscillation (CH1 and CH2). For the OEPO configuration, a pump signal at $f_p$ is supplied by a signal generator to the mixer's local oscillator (LO) port, whose phase can be tuned via $\varphi_p$. In the degenerate case, this pump at $f_p$ induces a auto-oscillating signal flowing inside the loop at $f_p/2$. The pump signal also supplies the reference trigger to the oscilloscope, branched out using a 10-dB coupler and then a frequency divider ($f\div 2$), which is also at $f_p/2$. Notably, inside the MZI, the positions of the phase-changing block (blue solid box) and the signal-sampling block (magenta dashed box) can be swapped, such that for each branch, the signal can be sampled either before or after phase-changing.}  
 \label{Fig_scheme}
\end{figure*}

To date, existing demonstrations have primarily focused on inline loop configurations with relatively rigid and globally distributed phase tunability \cite{xiong2020experimental,xiong2024magnon,wu2025coupling,barry2023ferrimagnetic,mi2026ultrasensitive,crescini2021phase}. Here, we introduce a Mach–Zehnder interferometer (MZI)-integrated magnonic OEO/OEPO architecture consisting of a magnonic branch and a reference phase-shifting branch. The key components in the magnonic branch are yttrium iron garnet (Y$_3$Fe$_5$O$_{12}$, YIG) \cite{serga2010yig} and the associated microwave stripline structure. 

We reveal that the introduced MZI degree of freedom results in novel spectral controllability and unconventional eigenmode characteristics in distributed auto-oscillating systems: (1) the MZI can partially soften the otherwise rigid parametric frequency pinning of the OEPO, leading to phase-sensitive frequency pulling \cite{fan2021injection} and transitional spectral adaptation even within a nominally ``fixed-frequency'' parametric oscillator; (2) placing the magnonic component inside one branch of the MZI, rather than inline along the main loop, substantially enhances the tunability and sensitivity of the phase and amplitude responses through coherent interferometric redistribution and regeneration; (3) quantitative complex-Lorentzian analysis further reveals phase-to-amplitude conversion and cross-branch coupling with distinct characters between the OEO and OEPO configurations. 

Our work reveals that the MZI serves not as a passive phase element but rather, as an active control knob in distributed, hybrid magnonic-photonic platforms. It establishes a new framework for interferometrically-engineered magnonic auto-oscillators and paves the way toward highly reconfigurable coherent magnonic-photonic platforms for sensing, signal processing, and classical-to-quantum hybrid information systems.

\section{Experiments}

Figure \ref{Fig_scheme} illustrates the experimental setup. The central build is an OEO loop that includes an electronic sector (consisting of microwave components for amplifying, attenuating, band-passing, and spectral detection) and a photonic sector (consisting of E-O/O-E converters and a fiber delay). For a proof-of-concept, the photonic part can be conveniently implemented by using fiber-optic patch cords and off-the-shelf RF-over-Fiber (RFoF) systems \cite{wu2025coupling}. On the right end, an electrical parametric pumping sector can be incorporated into the loop via the local oscillator (LO) port of a frequency mixer -- the nonlinear electrical device -- consisting of a signal generator supplying a pump signal, $f_p=2\pi\omega_p$, a tunable phase shifter ($\varphi_p$), and a 10-dB coupler to branch out the pump signal. This pumping sector transforms the system into an OEPO: the pump signal, at $f_p$, induces either a pair of nondegenerate frequencies, $f_{p1}=2\pi\omega_{p1}$ and $f_{p2}=2\pi\omega_{p2}$, or a degenerate, single frequency at half-pump, $f_p/2$, which we employ \cite{wu2025coupling}. In either case, the induced frequencies enter the OEO loop and form sustained auto-oscillations, which are subject to the amplification and phase-tuning of the main loop, rather than from the pump.    

A key distinction of our  setup from previous constructions in the literature is the central MZI structure at the microwave sector: the microwave signal first divides, at the splitter, into two branches (labeled 1 and 2) and re-converges at the combiner. Branch-1 is the reference branch, consisting mainly of a tunable phase shifter (PS), with a nominal phase set at $\varphi_\mathrm{ps}$, and an attenuator (for power balancing). Branch-2 is the magnonic branch, consisting mainly of the YIG-stripline structure with tunable external bias magnetic field (placed inside an electromagnet). Couplers are used at desirable locations in both branches, the main loop, and the pump line, to sample power (via power detectors, DI), spectrum (via spectrum analyzer, SA), and real-time oscillations (via oscilloscope, OSC). Using a frequency divider ($f \div2$) in the pump line, the phase of the two MZI branches and their difference can be determined by referencing to the global external pump signal -- used as the oscilloscope trigger. 

The loop signal is flowing one-way along the clockwise direction. Hence, at the bottom of the loop, the signal enters the MZI at the splitter and merging the two branches at the combiner (from right to left). Therefore, inside the MZI, the positions of the phase-changing components block (blue solid box) and the signal-sampling components block (magenta dashed box) can be swapped, such that for each branch, the signal can be sampled either before or after phase-changing.

\section{Results and Discussions}

\subsection{Determination of the cavity FSR and delay time}

With the added pump line, there are two fundamentally different ways to determine the FSR of the cavity loop.  

\begin{figure}[htb]
 \centering
 \includegraphics[width=3.4 in]{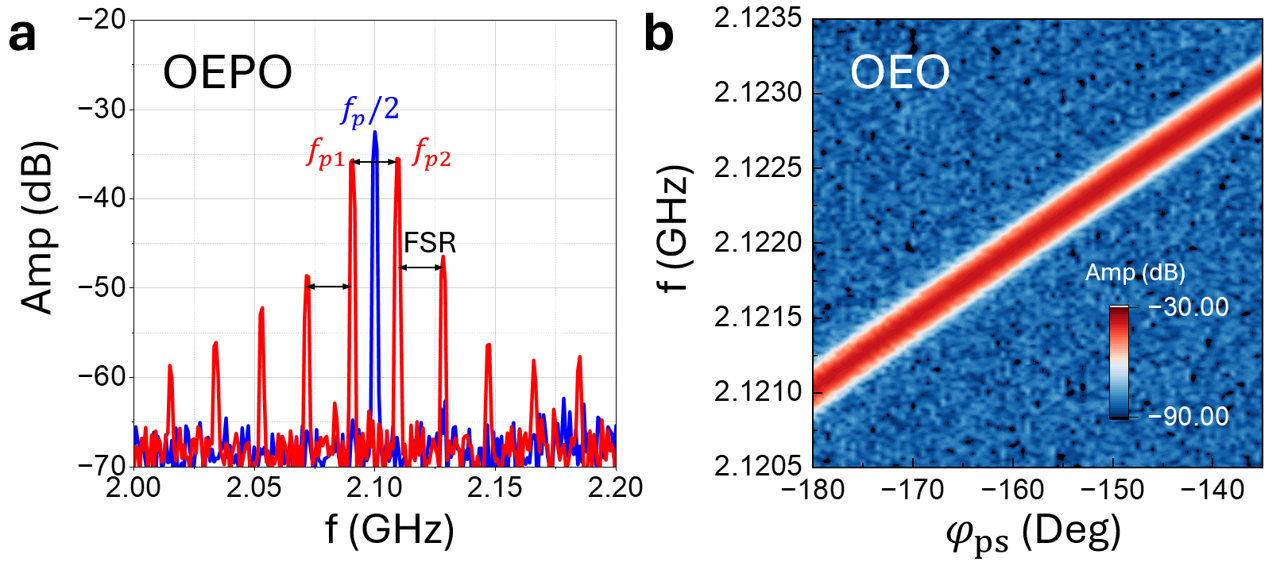}
 \caption{Two independent methods to determine the cavity FSR. (a) In the OEPO case, the FSR can be directly read out from the separation between the nondegenerate mode pairs, e.g., $f_\mathrm{p1}$ and $f_\mathrm{p2}$, yielding 18.2 MHz. The degenerate situation yields a single strong peak at $f_p/2$. The traces are obtained at Atten = 13 dB. Red peaks: $\varphi_\mathrm{ps} = -25^\circ$, Blue peak: $\varphi_\mathrm{ps} = -100^\circ$. (b) In the OEO case, the FSR is calculated by the frequency pulling, yielding a nominal value of 15.8 MHz and an effective $\beta_\mathrm{MZI} \approx 0.87$. The trace is obtained at Atten = 20 dB.    }  
 \label{Fig_FSR}
\end{figure}

In the OEPO case, by injecting a pump (e.g. $f_p = 4.2$ GHz), one can directly read out the FSR from the nondegenerate mode pairs , as in Fig.~\Ref{Fig_FSR}(a). The FSR is determined from the separation of red-colored peaks, FSR $\approx 18.2$ MHz, corresponding to a loop delay, $\tau_\mathrm{OEPO} \approx 54.9$ ns. Such a loop delay corresponds to a cavity length of $\approx 11.0$ m, which is consistent with the total cable length used in the experimental setup ($\sim6$-m optical fiber and $\sim 5$-m rf cable). On the other hand, in the OEO case, the FSR can also be estimated from the phase-induced frequency pulling of the loop's eigenmode. In our setup, by using appropriate band-pass filters, we can select a fundamental eigenmode of the OEO cavity near 2.12 GHz, as shown in Fig.~\ref{Fig_FSR}(b). The frequency of the OEO mode can be fine-tuned by the phase shifter, $\varphi_\mathrm{ps}$:
\vspace{-8pt}
\begin{align}
f(\varphi_\mathrm{ps}) = f_o + \frac{df}{d\varphi_\mathrm{ps}} \varphi_\mathrm{ps}, 
\end{align}
\vspace{-10pt}

\noindent in which $f_o$ is the frequency offset and $\frac{df}{d\varphi_\mathrm{ps}}$ is the frequency pulling slope (pulling coefficient). This slope allows to extract the loop delay, $\tau_\mathrm{OEO}$, via: 
\vspace{-8pt}
\begin{align}
\tau_\mathrm{OEO} \approx -\frac{\beta_\mathrm{MZI}}{2\pi} \left( \frac{df}{d\varphi_\mathrm{ps}} \right)^{-1}, 
\end{align}
\vspace{-10pt}

\noindent where $\beta_\mathrm{MZI}=\frac{d\varphi_\mathrm{MZI,eff}}{d\varphi_\mathrm{ps}}$ and is a coefficient unique to the present MZI structure, which accounts for the conversion of the local branch phase shift into an effective loop phase perturbation. This coefficient is usually close to but less than 1, i.e., $\beta_\mathrm{MZI} \lesssim 1$. From our result in Fig.~\ref{Fig_FSR}(b), the pulling coefficient, $\frac{df}{d\varphi_\mathrm{ps}} \approx 44$ kHz/Deg, which corresponds to a nominal cavity loop delay (assuming $\beta_\mathrm{MZI}=1$), $\tau_\mathrm{OEO} \approx 63$ ns, and an FSR $\approx 15.8$ MHz. Reconciling this value with the ground-truth FSR determined from the mode pairs in Fig.~\ref{Fig_FSR}(a) then yields an effective $\beta_\mathrm{MZI} \approx 0.87$.

\subsection{OEPO softening and frequency pulling}

\begin{figure*}[htb]
 \centering
 \includegraphics[width=5.6 in]{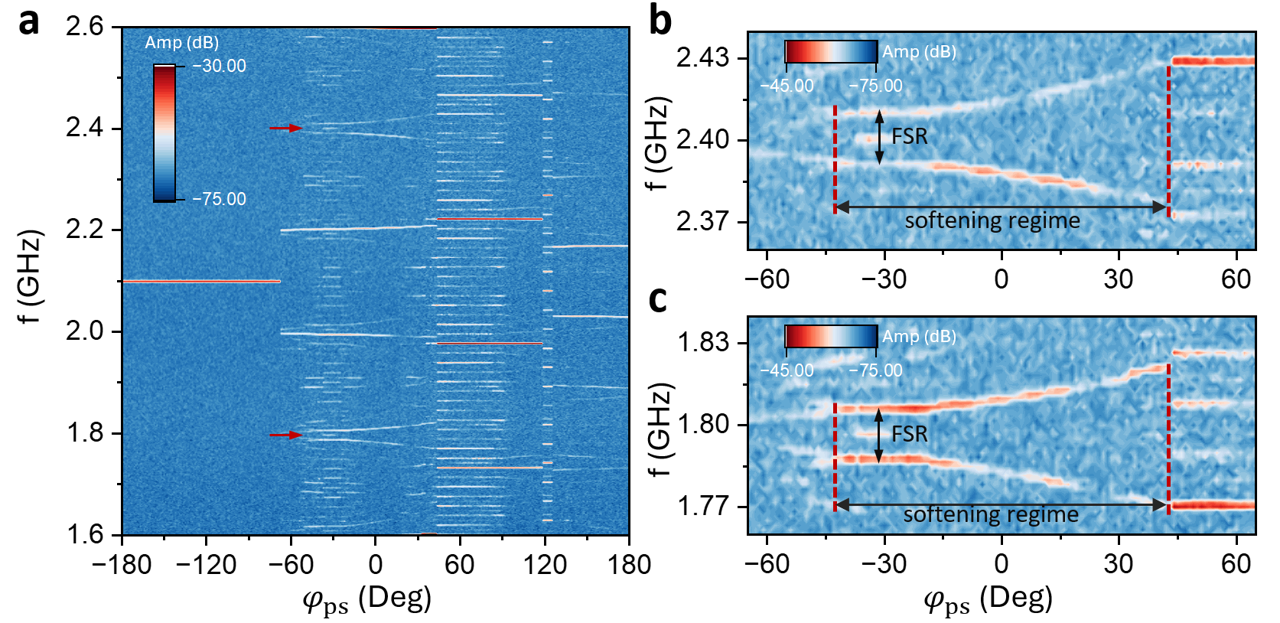}
 \caption{Spectrum of MZI-enhanced OEPO. (a) The full spectrum by scanning the $\varphi_\mathrm{ps}$ from $\varphi_\mathrm{ps} = -180^\circ$ to 180$^\circ$, under a pump $f_p=4.2$ GHz at 8 dBm power. The main loop is at Atten = 10.5 dB. (b,c) Zoom-in regime showing the frequency pulling near (b) 2.4 and (c) 1.8 GHz. The softening regime is identified near $\varphi_\mathrm{ps} \in (-43^\circ, 43^\circ)$, corresponding to an effective locking bandwidth of \(\Delta f_{\mathrm{lock}}=\Delta_{\mathrm{lock}}/(2\pi)\approx1.89~\mathrm{MHz}\).  }  
 \label{Fig_OEPO}
\end{figure*}

The key of MZI is the introduction of the modified transfer function, $H_\mathrm{MZI}$. For our system, branch-1 corresponds to the reference phase-shifter branch, while branch-2 corresponds to the magnonic YIG branch. The total MZI transfer function can therefore be written as:
\vspace{-6pt}
\begin{align}
H_{\mathrm{MZI}}(\omega,H,\varphi_{\mathrm{ps}})
&=
A_{\mathrm{ps}}
e^{i(\omega\tau_1+\varphi_{\mathrm{ps}})} 
\nonumber\\
&\quad+
A_{\mathrm{YIG}}(\omega,H)
e^{i[\omega\tau_2+\varphi_{\mathrm{YIG}}(\omega,H)]}, 
\label{Eq_MZI}
\end{align}
\vspace{-18pt}

\noindent where \(A_{\mathrm{ps}}\) is the effective amplitude of the phase-shifter branch, \(\tau_1\) is the delay of the phase-shifter branch, \(\varphi_{\mathrm{ps}}\) is the externally controlled phase shift, \(A_{\mathrm{YIG}}(\omega,H)\) is the resonant amplitude response of the YIG branch, \(\tau_2\) is the delay of the YIG branch, and \(\varphi_{\mathrm{YIG}}(\omega,H)\) is the resonant dispersive phase response of the YIG branch. In our case, we adjusted the cable lengths for the two branches so that $\tau_1 \approx \tau_2 \approx \tau$, thus:
\vspace{-6pt}
\begin{align}
H_{\mathrm{MZI}}
&=
e^{i\omega\tau}
\left[
A_{\mathrm{ps}}e^{i\varphi_{\mathrm{ps}}}
+
A_{\mathrm{YIG}}(\omega,H)
e^{i\varphi_{\mathrm{YIG}}(\omega,H)}
\right].
\label{Eq_MZI_simple}
\end{align}
\vspace{-18pt}

The total transfer function described in Eq.~\ref{Eq_MZI_simple} makes the system significantly more prone to softening of the OEPO modes. As a result, in certain regimes where the OEPO condition (phase and gain) is not met, frequency pulling (a hallmark of an OEO) can arise due to the sensitive MZI phase tunability. 

The MZI-induced effective loop phase is determined by coherent vector recombination through: 
\vspace{-6pt}
\begin{align}
H_{\mathrm{MZI}}=|H_{\mathrm{MZI}}|e^{i\theta_{\mathrm{MZI}}}, \theta_{\mathrm{MZI}}=\arg(H_{\mathrm{MZI}}),
\end{align}
\vspace{-18pt} 

\noindent where $\theta_{\mathrm{MZI}}$ is the effective MZI phase experienced by the oscillator loop. Consequently, the effective phase is not a simple additive quantity as in the inline counterpart \cite{wu2025coupling}, but is instead highly sensitive to branch imbalance and interferometric conditions. 

With the present MZI, the effective phase mismatch can be defined: 
\vspace{-8pt}
\begin{align}
\Delta_{\mathrm{eff}}=2\omega-\omega_p+\frac{1}{\tau_{\mathrm{eff}}}\theta_{\mathrm{MZI}}(\omega,\varphi_\mathrm{ps}),
\label{Eq_mismatch} 
\end{align}
\vspace{-12pt}

\noindent where \(\tau_{\mathrm{eff}}\) is the effective loop delay, \(\theta_{\mathrm{MZI}}\) acts as an additional phase-dependent detuning term. A stable degenerate state requires: $|\Delta_{\mathrm{eff}}|<\Delta_{\mathrm{lock}}$, where \(\Delta_{\mathrm{lock}}\) is the effective parametric locking bandwidth. 

The MZI phase response varies with both frequency and applied phase: $\delta\theta_{\mathrm{MZI}}=\frac{\partial \theta_{\mathrm{MZI}}}{\partial \varphi_\mathrm{ps}}\delta\varphi_\mathrm{ps}+\frac{\partial \theta_{\mathrm{MZI}}}{\partial \omega}\delta\omega$, where the first term is dominant in the present situation (externally change $\varphi_\mathrm{ps}$ via phase shifter). Then, the effective detuning becomes approximately:
\vspace{-6pt}
\begin{align}
\Delta_{\mathrm{eff}} \approx 2\delta\omega + \frac{\beta_{\mathrm{MZI}}}{\tau_{\mathrm{eff}}} \delta\varphi_\mathrm{ps}.
\label{Eq_mismatch_simple}
\end{align}
\vspace{-12pt}

In the rigid degenerate OEPO regime, the oscillator enforces $\delta\omega \approx 0$ and absorbs perturbations through coherent phase redistribution, gain redistribution, and amplitude adaptation. However, if the MZI-induced perturbation becomes sufficiently large: $\left| \frac{\beta_{\mathrm{MZI}}\delta\varphi_\mathrm{ps}}{\tau_{\mathrm{eff}}} \right| \gtrsim \Delta_{\mathrm{lock}}$, then the degenerate state can no longer remain perfectly pinned. The oscillator then partially relaxes the phase mismatch through frequency pulling: $\delta\omega \approx - \frac{\beta_{\mathrm{MZI}}}{2\tau_{\mathrm{eff}}} \delta\varphi_\mathrm{ps}$. Under such conditions, the otherwise rigid degenerate OEPO state becomes softened, allowing partial spectral adaptation and observable frequency  \cite{razavi2004study,markovic2019injection,li2026magnonic}. This explains why frequency-pulling behavior and softened parametric locking are much more readily observed in the MZI-OEPO configuration than in conventional inline series architectures. In Appendix A, as a comparison, we describe the transfer function and the associated response for the inline counterpart.  

We can also experimentally estimate the $\Delta_{\mathrm{lock}}$ from the transition boundary between rigidly pinned OEPO behavior and softened frequency-pulling behavior -- the onset of softening approximately occurs when the effective interferometric mismatch satisfies \(|\Delta_{\mathrm{eff}}|\sim\Delta_{\mathrm{lock}}\). For the MZI-integrated system, the effective mismatch near the transition can be approximated as \(\Delta_{\mathrm{eff}}\sim \beta_{\mathrm{MZI}}\varphi_\mathrm{ps, crit}/\tau_{\mathrm{eff}}\), where \(\varphi_{\mathrm{ps,crit}}\) corresponds to the critical $\varphi_\mathrm{ps}$ value at which the OEPO first begins to exhibit observable frequency pulling. 

From Fig.~\ref{Fig_OEPO}(b,c), we identify that the softening regime occurs roughly from $\varphi_\mathrm{ps} \sim -43^\circ$ to $43^\circ$. Using \(\varphi_{\mathrm{ps,crit}}=43^\circ\), \(\tau_{\mathrm{eff}}=\tau_{\mathrm{OEPO}}\sim54.9~\mathrm{ns}\) with \(\beta_{\mathrm{MZI}}\sim 0.87\), one obtains \(\Delta_{\mathrm{lock}}\approx0.75\times 0.87/(54.9~\mathrm{ns})\approx1.187\times10^7~\mathrm{rad/s}\), corresponding to an effective locking bandwidth of \(\Delta f_{\mathrm{lock}}=\Delta_{\mathrm{lock}}/(2\pi)\approx1.89~\mathrm{MHz}\). This value is reasonable for the present system, as it is substantially smaller than the experimentally observed loop \(\mathrm{FSR}\sim18~\mathrm{MHz}\), allowing the OEPO to remain pinned within a single mode without hopping into adjacent loop modes. At the same time, the locking bandwidth is sufficiently small that moderate interferometric phase perturbations introduced by the MZI can become comparable to the locking stiffness, thereby enabling observable mode softening and partial frequency pulling. Quantitatively, the ratio \(\Delta f_{\mathrm{lock}}/\mathrm{FSR}\sim1.89/18\approx0.1\) indicates that the locking bandwidth corresponds to only approximately \(10\%\) of the loop mode spacing. Physically, this suggests that the OEPO remains strongly frequency pinned, but not infinitely rigid, such that the MZI-induced interferometric mismatch can become sufficiently large to partially overcome the parametric restoring force before the system transitions into nondegenerate mode-pair states. 

\subsection{Adaptive interferometric oscillation} 

One key attribute of the present MZI architecture compared to conventional parametric oscillator is the ability to adaptively and coherently distribute (and redistribute) local perturbations via an interferometric approach. In other words, each branch no longer responds to any externally imposed perturbation purely locally, but partially compensates and redistributes the phase and amplitude through coherent feedback and interferometric adaptation. Such adaptive behaviors become directly evident in the experimentally-observed realtime oscillations (captured via the oscilloscope). In particular, by sampling each branch's signals before and after their respective phase-changing elements, one can distinguish the globally redistributed adaptive oscillations (signal-sampling $\rightarrow$ phase-changing) and the direct local phase accumulation introduced by each phase-changing element itself (phase-changing $\rightarrow$ signal-sampling). Thus, the comparison between the two sampling locations provides a powerful means to separate collective coherent redistribution from local branch-specific phase evolution within the adaptive MZI oscillator.

\begin{figure}[htb]
 \centering
 \includegraphics[width=3.2 in]{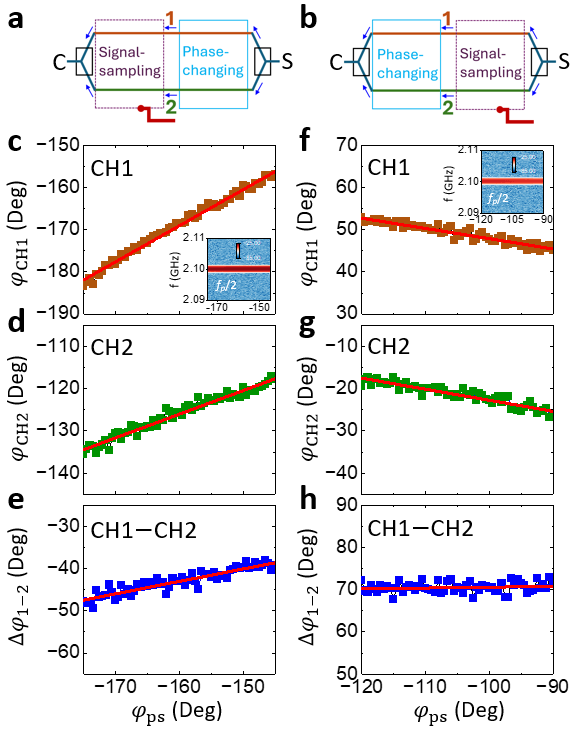}
 \caption{(a,b) Two sampling positions in which: (a) signal-sampling after the phase-changing elements (phase-changing $\rightarrow$ signal-sampling), and (b) signal-sampling before the phase-changing elements (signal-sampling $\rightarrow$ phase-changing). C: combiner, S: splitter. (c,d,e) phase evolution as a function of the phase-shifter nominal set phase, $\varphi_\mathrm{ps}$,  for the case in (a) of (c) phase shifter branch (CH1), (d) YIG branch (CH2), and (e) the net phase. (f,g,h) phase evolution as a function of the phase-shifter nominal set phase, $\varphi_\mathrm{ps}$, for the case in (b) of (g) phase shifter branch (CH1), (f) YIG branch (CH2), and (h) the net phase. Inset of (c) and (f): OEPO spectrum showing the stable $f_p/2$ mode throughout the set phase range for the respective sampling positions. The measurement and phase spectrum were taken at: (c-e) Atten = 11 dB. (f-h) Atten = 15.5 dB. }  
 \label{Fig_PS}
\end{figure}

For this purpose, we focus on the OEPO's degenerate $f_p/2 (= 2.1 ~\mathrm{GHz})$ mode and select appropriate $\varphi_\mathrm{ps}$ windows for the two sampling positions in Fig.~\ref{Fig_PS}(a): phase-changing $\rightarrow$ signal-sampling, and Fig.~\ref{Fig_PS}(b): signal-sampling $\rightarrow$ phase-changing. An incoming wave entering the splitter (right-end) can be expressed as: 
\vspace{-8pt}
\begin{align}
V_{\mathrm{in}}=A_{\mathrm{in}}e^{i\varphi_{\mathrm{global}}},
\label{Eq_Vin}
\end{align} 
\vspace{-18pt}

\noindent where $\varphi_\mathrm{global}$ is the adaptive, globally re-distributed phase, and $A_\mathrm{in}$ is the amplitude. For OEPO, the fixed wave frequency at $f_p/2$ simplifies the situation, as shown by the inset figures in Fig.~\ref{Fig_PS}(c,f). After the phase shifter and at the combiner (left-end), this wave becomes: 
\vspace{-8pt}
\begin{align}
V_{\mathrm{out}}=A_{\mathrm{in}}e^{i(\varphi_{\mathrm{global}}+\varphi_{\mathrm{local}})}.
\label{Eq_Vout}
\end{align} 
\vspace{-18pt}

\noindent Consequently, the case in Fig.~\ref{Fig_PS}(a) measures $\varphi_{\mathrm{global}}+\varphi_{\mathrm{local}}$ and the case in Fig.~\ref{Fig_PS}(b) measures only $\varphi_{\mathrm{global}}$. Since the oscillator adapts globally to maintain the loop phase condition, both the global and local phases are functions of the externally imposed phase, namely \(\varphi_{\mathrm{global,local}}=\varphi_{\mathrm{global,local}}(\varphi_{\mathrm{ps}})\). Therefore, sweeping the $\varphi_\mathrm{ps}$ value within the appropriate range (that stabilizes the $f_p/2$ mode) allows extracting the phase coefficients for both branches before and after the phase shifter, theoretically -- before (near splitter): $s_\mathrm{splitter}= d\varphi_\mathrm{global}/d\varphi_\mathrm{ps}$; and after (near combiner): $s_\mathrm{combiner}=(d\varphi_\mathrm{global}+d\varphi_\mathrm{local})/d\varphi_\mathrm{ps} = s_\mathrm{splitter}+d\varphi_\mathrm{local}/d\varphi_\mathrm{ps}$.    

Figure~\ref{Fig_PS}(c-e) show the measured phase for the phase-shifter branch (CH1), the YIG branch (CH2), and the net phase (CH1$-$CH2), for the sampling configuration in Fig.~\ref{Fig_PS}(a), and Fig. ~\ref{Fig_PS}(f-h) show the measured phase for the respective branches and the net, for the sampling configuration in Fig.~\ref{Fig_PS}(b). All the fitted phase coefficients are summarized in Table ~\ref{tab:phase_slopes}.

The slope in Fig.~\ref{Fig_PS}(f,g) is negative, i.e., \(d\varphi_{\mathrm{global}}/d\varphi_{\mathrm{ps}}<0\), indicating that the oscillator globally re-distributes the phase in the opposite direction of the externally imposed phase perturbation $\varphi_\mathrm{ps}$. Physically, this corresponds to adaptive negative-feedback phase compensation inside the coherent oscillator loop. In addition, the re-distributed $\varphi_\mathrm{global}$ equally impacts both branches (CH1 and CH2), i.e., sharing a similar negative global compensation, and hence, yielding a net phase coefficient $\sim0$, as shown in Fig.~\ref{Fig_PS}(h).   

On the other hand, the slope in  Fig.~\ref{Fig_PS}(c,d) is positive, because the directly imposed local phase contribution still outweighs the compensating global redistribution. Notably, while the phase-shifter branch contains the directly imposed local phase perturbation, in Fig.~\ref{Fig_PS}(c), the YIG branch receives a finite induced phase contribution through adaptive interferometric redistribution inside the coherent oscillator loop. The coefficient for the YIG branch is less than that of the phase shifter branch (Fig.~\ref{Fig_PS}(d)), yielding a net phase coefficient of $\sim0.295$, in Fig.~\ref{Fig_PS}(e). 

\vspace{-12pt}
\begin{table}[htb]
\centering
\caption{Measured phase slopes under phase-shifter tuning for different sampling configurations.}
\begin{tabular}{c||c|c}
\hline
 & $s_\mathrm{combiner}$ [Fig.~\ref{Fig_PS}(a)] & $s_\mathrm{splitter}$ [Fig.~\ref{Fig_PS}(b)] \\
\hline
coefficient 
& $\dfrac{d\varphi_{\mathrm{global}}}{d\varphi_{\mathrm{ps}}}+\dfrac{d\varphi_{\mathrm{local}}}{d\varphi_{\mathrm{ps}}}$ 
& $\dfrac{d\varphi_{\mathrm{global}}}{d\varphi_{\mathrm{ps}}}$ \\
\hline
CH1 & $0.854\pm0.011$ & $-0.243\pm0.011$ \\
CH2 & $0.560\pm0.014$ & $-0.262\pm0.013$ \\
CH1$-$CH2 & $0.295\pm0.015$ & $\sim 0$ \\
\hline
\end{tabular}
\label{tab:phase_slopes}
\end{table}
\vspace{-8pt}

Last but not least, the difference of the before and after slopes, $s_\mathrm{combiner}-s_\mathrm{splitter} = \frac{d\varphi_{\mathrm{local}}}{d\varphi_{\mathrm{ps}}}$ allows to directly examine the phase effect within each branch. For CH1, $\frac{d\varphi_{\mathrm{local, CH1}}}{d\varphi_{\mathrm{ps}}} \sim 1.1$, and $\frac{d\varphi_{\mathrm{local, CH2}}}{d\varphi_{\mathrm{ps}}} \sim 0.82$. Therefore, the YIG branch picks up about \(0.82/1.10\approx0.75\) of the phase-shifter branch response. The experimentally observed coexistence of negative global phase response and positive local phase response constitutes strong evidence for distributed phase self-compensation and adaptive coherent redistribution inside the interferometric auto-oscillating system.

\subsection{YIG resonance and dispersive phase shift}

A conventional phase shifter as discussed in the earlier section approximately produces, as in Eq.~\ref{Eq_Vin} and Eq.~\ref{Eq_Vout}:  $V_{\mathrm{out}}=A_{\mathrm{in}}e^{i(\varphi_{\mathrm{global}}+\varphi_{\mathrm{local}})}=V_{\mathrm{in}}e^{i\varphi_{\mathrm{local}}(\varphi_\mathrm{ps})}$, where the $\varphi_\mathrm{ps}$ is the externally imposed phase and the amplitude remains nearly unchanged. Such a response is broadband and nonresonant. Sweeping the phase shifter, therefore, directly modifies the global interferometric boundary condition of the MZI. However, the YIG branch behaves fundamentally differently in that the YIG-induced phase is not externally imposed but instead originates from the resonant dispersive susceptibility of the magnonic system. Near ferromagnetic resonance (FMR), the YIG transfer function can be described as: 
\vspace{-8pt}
\begin{align}
H_{\mathrm{YIG}}(\omega,H)
=
A_\mathrm{YIG}(\omega,H)e^{i\varphi(\omega,H)}
\label{YIG_function} 
\end{align}
\vspace{-18pt}

\noindent where both the amplitude, $A_\mathrm{YIG}$ and the phase, $\varphi_\mathrm{YIG}(\omega,H)$, vary strongly near resonance due to the YIG's complex susceptibility: \(\chi(H)=\chi'(H)+i\chi''(H)\), where \(\chi''\) corresponds to absorption loss while \(\chi'\) corresponds to dispersive phase rotation and delay. 

\begin{figure}[htb]
 \centering
 \includegraphics[width=3.1 in]{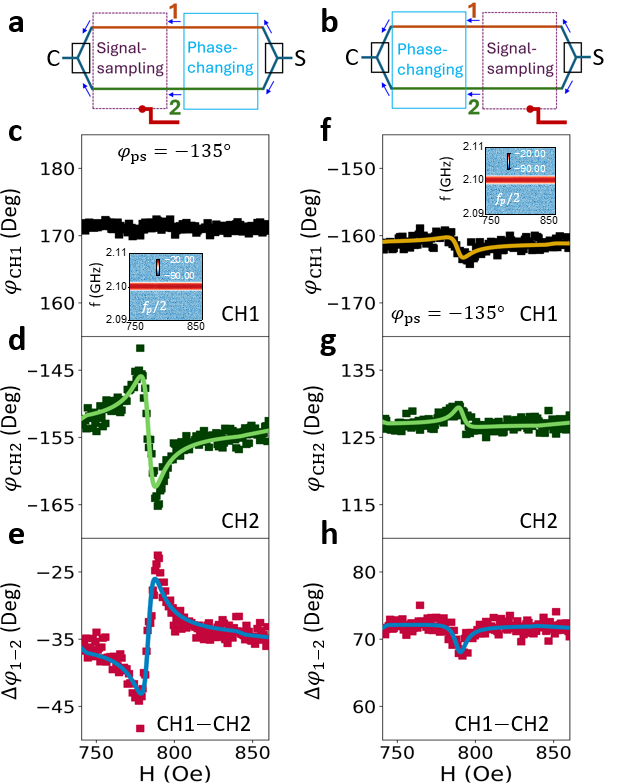}
 \caption{(a,b) Two sampling positions in which: (a) phase-changing $\rightarrow$ signal-sampling, and (b) signal-sampling $\rightarrow$ phase-changing. C: combiner, S: splitter. (c,d,e) phase evolution as a function of the magnetic field, $H$, for the case in (a) of (c) phase shifter branch (CH1), (d) YIG branch (CH2), and (e) the net phase. (f,g,h) phase evolution as a function of the magnetic field, $H$, for the case in (b) of (f) phase shifter branch (CH1), (g) YIG branch (CH2), and (h) the net phase. Inset of (c) and (f): OEPO spectrum showing the stable $f_p/2$ mode throughout the magnetic field range for the respective sampling positions. The measurement and phase spectrum were taken at: (c-e) Atten = 10.5 dB. (f-h) Atten = 10.5 dB.  }  
 \label{Fig_YIG}
\end{figure}

Figure~\ref{Fig_YIG} shows the YIG-induced phase and amplitude perturbations to both branches for the two respective sampling configurations, Fig.~\ref{Fig_YIG}(a,b). First, when the signals are sampled near the combiner (after YIG and phase shifter), the phase of the phase shifter branch is insensitive to YIG-induced perturbation, Fig.~\ref{Fig_YIG}(c). This is because the YIG perturbation is predominantly dispersive, thus the induced perturbation remains largely localized within the YIG branch, without direct impact to the phase shifter branch (CH1). In stark contrast, sampling at the YIG branch itself (CH2) shows a large dispersive phase response, Fig.~\ref{Fig_YIG}(d), and hence the net (CH1-CH2), Fig.~\ref{Fig_YIG}(e). This dispersive phase character is physically consistent with the resonant nature of the YIG susceptibility. In this sampling configuration, the measured signal contains the full local resonant phase accumulation \(\delta\varphi_{\mathrm{YIG}}(H)\) induced by the FMR. 

On the other hand, when signal-sampling occurs before phase-changing, after coherent regeneration through the feedback loop, the redistributed global oscillation contains the YIG-induced phase perturbation, allowing upstream sampling near the splitter, Fig.~\ref{Fig_YIG}(b), to reveal measurable phase response in both branches (CH1 and CH2), in Fig.~\ref{Fig_YIG}(f,g), respectively. The amount of phase variation for CH1 and CH2 are similar, indicating that both branches inherit similar overall YIG-induced perturbation amplitudes and exhibit comparable mixed absorptive-dispersive Lorentzian characteristics. However, their slightly different lineshapes still result in a net phase change (CH1-CH2), in Fig.~\ref{Fig_YIG}(h). Because the two branches remain physically distinct in propagation paths with different local transfer functions, the interferometric projection weights into the regenerated oscillation. Therefore, although the globally redistributed perturbation becomes common to both branches, it does not project onto the two branches with exactly the same complex phase composition. Consequently, the residual Lorentzian-phase mismatch effectively quantifies how differently the global adaptive oscillation projects back onto each interferometric branch.

The field-dependent responses measured from the phase-shifter branch (CH1), the YIG branch (CH2), and their net difference (CH1$-$CH2), shown in Fig.~\ref{Fig_YIG}, may be described using mixed symmetric--antisymmetric Lorentzian functions. The detailed model is described in Appendix B. Fitting the responses to Lorentzian functions yields resonance amplitudes (\(A_{\mathrm{ps}}\), \(A_{\mathrm{YIG}}\), \(A_{\mathrm{diff}}\)) and Lorentzian phases (\(\psi_{\mathrm{ps}}\), \(\psi_{\mathrm{YIG}}\), \(\psi_{\mathrm{diff}}\)) that determine the relative weights of the symmetric absorptive and antisymmetric dispersive components. The fitted parameters are summarized in Table ~\ref{tab:lorentzian_parameters}. 

\begin{table}[htb]
\centering
\caption{Fitted complex Lorentzian parameters for different sampling configurations.}
\begin{tabular}{c|cc|cc}
\hline
& \multicolumn{2}{c|}{After [Fig.~\ref{Fig_YIG}(a)]} & \multicolumn{2}{c}{Before [Fig.~\ref{Fig_YIG}(b)]} \\
\cline{2-5}
& $A$ & $\psi$ & $A$ & $\psi$ \\
\hline
CH1 (ps) & $\sim 0$ & NA & 3.0 & $-117.5$ \\
CH2 (YIG) & 16.5 & $-93.8$ & 3.0 & $-40.2$ \\
CH1$-$CH2 (diff) & 17.1 & 85.9 & 4.1 & $-162.2$ \\
\hline
\end{tabular}
\label{tab:lorentzian_parameters}
\end{table}

Consequently, \(A_{\mathrm{diff}}\) quantifies the strength of the residual branch imbalance, while \(\psi_{\mathrm{diff}}\) characterizes the physical nature of the adaptive redistribution between the two branches. When the signals are sampled after the YIG, the differential Lorentzian phase is \(\psi_{\mathrm{diff}}=85.9^\circ\approx90^\circ\), Fig.~\ref{Fig_YIG}(e), therefore the response is predominantly antisymmetric and dispersive-like, indicating that the branch imbalance is dominated by phase redistribution. By contrast, when the sampling occurs before the YIG, both CH1 and CH2 measure intermediate Lorentzian phases, i.e. $\psi_{\mathrm{ps}}=-117.5^\circ$, Fig.~\ref{Fig_YIG}(f), and $\psi_{\mathrm{YIG}}=-40.2^\circ$, Fig.~\ref{Fig_YIG}(g), indicating a mixed amplitude and phase re-distribution after the loop's concerted phase and gain adjustment and eigenmode regeneration. In other words, both absorptive and dispersive imbalances coexist. Last but not least, the differential Lorentz phase measures \(\psi_{\mathrm{diff}}=-162.2^\circ\), Fig.~\ref{Fig_YIG}(h), suggesting that the two interferometric branches redistribute the YIG perturbation differently through a combined phase evolution and amplitude redistribution. An extended analytical model describing the adaptive phase-to-amplitude interconversion in the interferometric loop is outlined in Appendix B.   

\subsection{Global phase control vs. local phase ``defect''}

After elucidating the loop's adaptive response to phase perturbations of respective MZI branches, we turn to their mutual interactions by monitoring the phase response of one branch subjected to a ``phase bias'' from the other. Physically, the phase shifter branch primarily acts as a global interferometric boundary-control that governs how the adaptive oscillator redistributes externally imposed phase perturbations, whereas the YIG branch primarily modifies the local branch susceptibility and resonant phase variation. 

Experimentally, the local YIG Lorentzian response remains comparatively insensitive to the phase-shifter bias, as discussed in Appendix C. This corroborates that the YIG branch behaves predominantly as a local resonant dispersive subsystem. However, the measured phase slopes with respect to \(\varphi_{\mathrm{ps}}\) shows a nontrivial dependence over the YIG resonance condition induced by the magnetic field \(H\). 

\begin{figure*}[htb]
 \centering
 \includegraphics[width=5.6 in]{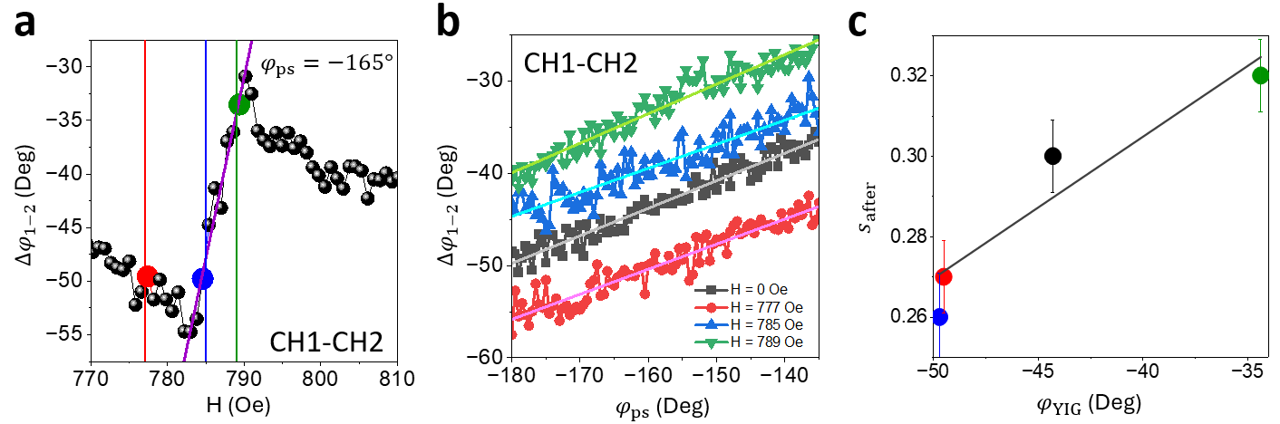}
 \caption{(a) CH1-CH2 signal acquired by sweeping $H$ at $\varphi_\mathrm{ps} = -165^\circ$. The slope is $\sim3.38$ Deg/Oe, corresponding to the coefficient of YIG-induced phase conversion. The critical data points corresponding to the selected $H$-field bias values are zoomed-in and color-coded. (b) CH1-CH2 signal acquired by sweeping $\varphi_\mathrm{ps}$ at different $H$ values, hence resulting in different $\varphi_\mathrm{YIG}(H)$. Slopes ($s_\mathrm{combiner}$): $H=0$ Oe, $s=0.30 \pm0.01$, $H=777$ Oe, $s=0.27\pm0.01$, $H=785$ Oe, $s=0.26\pm0.01$, $H=789$ Oe, $s=0.32\pm0.01$. (c) Fitting the $s_\mathrm{combiner}$ values to the $\varphi_\mathrm{YIG}(H)$ acquired at different $H$ bias. The slope is the cross-coupling coefficient \(\kappa_{\mathrm{YIG}\rightarrow\mathrm{ps}} \sim0.004 ~\mathrm{Deg}^{-1}\) .  }  
 \label{Fig_PS_slope}
\end{figure*}

We measured the phase shifter effects at different fields along the YIG resonance lineshape, see Fig.~\ref{Fig_PS_slope}. We first sweep the YIG resonance at a selected $\varphi_\mathrm{ps}=-165^\circ$ and measure the differential phase signal (CH1-CH2), Fig.~\ref{Fig_PS_slope}(a). Next, we selected representative $H$ values (0, 777, 785, 789 Oe) that exemplify different YIG resonance locations (hence the YIG-branch phase, $\varphi_\mathrm{YIG}(H)$) and measured the phase shifter branch susceptibility by scanning $\varphi_\mathrm{ps}$ from $-180^\circ$ to $-135^\circ$, see Fig.~\ref{Fig_PS_slope}(b). This is the $s_\mathrm{combiner}$ configuration (sampling near the combiner). The PS slopes measured for different fields is then plotted against the YIG phase $\varphi_\mathrm{YIG}$, see Fig.~\ref{Fig_PS_slope}(c), yielding a small but nontrivial dependence, suggesting a weak cross-branch coupling.     

This weak cross-coupling between the two branches arises through adaptive redistribution of the interferometric operating point. Specifically, we can rewrite the measured phase-shifter slope as: 
\vspace{-8pt}
\begin{align}
S_{\mathrm{ps}}(H)
=
S_0
+
\delta S_{\mathrm{YIG}}(H),
\end{align}
\vspace{-16pt}

\noindent to incorporate the YIG-induced correction, \(\delta S_{\mathrm{YIG}}(H)\), atop the baseline structural phase-redistribution slope, \(S_0\). More generally, this correction can come from either amplitude-mediated coupling or phase-mediated coupling: 
\vspace{-8pt}
\begin{align}
\delta S_{\mathrm{YIG}}(H)
&\propto
\frac{
\partial S_{\mathrm{ps}}
}{
\partial A_{\mathrm{YIG}}
}
\delta A_{\mathrm{YIG}}(H)
\nonumber\\
&\qquad
+
\frac{
\partial S_{\mathrm{ps}}
}{
\partial\varphi_{\mathrm{YIG}}
}
\delta\varphi_{\mathrm{YIG}}(H).
\end{align}
\vspace{-12pt}

The first term (amplitude-mediated coupling) represents the change in the branch balance caused by the YIG's resonant loss, so the PS phase perturbation is projected differently onto the recombined MZI eigenstate. The second term (phase-mediated coupling) dictates the change in the MZI operating point caused by the YIG phase shift, so the PS slope changes because the interferometer is biased at a slightly different effective relative phase. The latter term dominates our present experiment.

Experimentally, plotting the measured PS slope versus the YIG-induced phase reveals an approximately linear relation, as in Fig.~\ref{Fig_PS_slope}(c), therefore, 
\vspace{-6pt}
\begin{align}
S_{\mathrm{ps}}
=
S_0
+
\kappa_{\mathrm{YIG}\rightarrow\mathrm{ps}}
\varphi_{\mathrm{YIG}},
\end{align}
\vspace{-14pt}

\noindent where \(\kappa_{\mathrm{YIG}\rightarrow\mathrm{ps}}\) serves as a quantitative cross-susceptibility coefficient describing weak adaptive coupling between the two interferometric branches. Our fitted results yields a value of \(\kappa_{\mathrm{YIG}\rightarrow\mathrm{ps}} \sim0.004 ~\mathrm{Deg}^{-1}\). Therefore, the YIG resonance shifts the operating point of the oscillator on the adaptive phase manifold, while the PS sweep probes the local tangent of that manifold, which reveals a nontrivial dependence to the YIG phase bias. Consequently, the MZI-OEPO behaves neither as two fully independent resonators nor as a perfectly rigid common-mode oscillator, but rather as a partially synchronized distributed adaptive interferometric eigenstate with weak branch-dependent cross-coupling. 

\section{Summary}

In this work, we constructed and investigated a unique MZI-based optoelectronic magnonic parametric oscillator (OEMPO) incorporating a YIG-loaded magnonic branch and a tunable phase-shifter branch, enabling systematic exploration of adaptive interferometric oscillator dynamics under distributed phase perturbations. Through analysis of nondegenerate OEPO mode pairs and frequency-pulling behavior, the loop FSR and effective delay time were quantitatively extracted. Despite the nominally frequency-pinned parametric operation, we observed the emergence of OEPO mode softening and weak frequency pulling, indicating that the MZI geometry introduces an additional adaptive interferometric degree of freedom beyond conventional OEPO dynamics.

By comparing local and global sampling configurations, we further demonstrated that the YIG branch and the phase-shifter branch exhibit qualitatively distinct physical responses. The YIG subsystem behaves predominantly as a local dispersive resonant element governed by the complex magnonic susceptibility, whereas the phase-shifter branch primarily controls the global interferometric redistribution geometry. Nevertheless, coherent recombination and adaptive regeneration within the MZI loop produce finite cross-coupling between the two branches, allowing the globally regenerated oscillator eigenstate to inherit perturbations originating from either branch. Quantitative Lorentzian analysis further revealed substantial phase-to-amplitude conversion, branch-dependent adaptive redistribution, and partially synchronized interferometric dynamics. Comparison between the OEO and OEPO regimes additionally showed that the phase-pinned OEPO favors strongly dispersive local YIG response, whereas the frequency-adaptive OEO partially relaxes perturbations through spectral motion and therefore exhibits comparatively more mixed and absorptive local response.

More broadly, the present platform establishes a versatile framework for studying adaptive nonlinear interferometric physics in hybrid magnonic-photonic oscillator systems. The coexistence of local resonant susceptibility dynamics, global coherent redistribution, nontrivial branch-level cross-coupling, and interferometric amplitude-phase conversion suggests rich opportunities for future exploration of branch-dependent synchronization phenomena, adaptive nonlinear phase redistribution, coherent delay engineering, and phase-sensitive magnonic information processing. Beyond YIG-based systems, the underlying architecture may also be extended toward other resonant quantum and wave-based platforms, including superconducting resonators \cite{trevillian2026temporal,kronowetter2023quantum,rani2025high}, metasurfaces \cite{xiong2024combinatorial,xiong2025photon,xiong2024hybrid}, photonic-magnonic integrated circuits \cite{zhu2020waveguide,pintus2025integrated,huang2017integrated,liang2023chip}, and emerging low-dimensional magnetic materials \cite{tang2023spin}, thereby opening new directions for adaptive coherent oscillator networks and distributed interferometric nonlinear dynamics.

\section{Acknowledgments} 

The experimental work at UNC-CH was supported by the U.S. Department of Energy, Office of Science, Basic Energy Sciences under Award Number DE-SC0026305. Q. Gu acknowledges support from the National Science Foundation under Award Number ExpandQISE-2329027.

\section{Appendices}

\subsection{A. Inline phase shifter transfer function}
\label{app:series}

\renewcommand{\theequation}{A-\arabic{equation}}
\setcounter{equation}{0}  
\renewcommand{\thefigure}{A-\arabic{figure}}
\setcounter{figure}{0}  

Compared to the MZI configuration discussed in the present work, the inline series configuration is less prone to generating a sufficiently large effective mismatch because the perturbation enters the loop as a relatively smooth and uniform phase offset, rather than through highly sensitive interferometric recombination. For the inline case, where the two components are in series:
\vspace{-8pt}
\begin{align}
H_{\mathrm{ser}}(\omega,H,\varphi_{\mathrm{ps}})
&=
A_\mathrm{ps} A_{\mathrm{YIG}}(\omega,H)
\nonumber\\
&\quad\times
e^{i[\omega\tau+\varphi_{\mathrm{ps}}
+\varphi_{\mathrm{YIG}}(\omega,H)]},
\label{Eq_Ser}
\end{align}
\vspace{-18pt}

\noindent in which case the phases are all lumped and thus lack the interferometric features. As a result, the inline OEPO modes are largely rigid, showing mostly mode hopping once power or phase changes the loop condition. For the inline series case, the total transfer function is in a form such that the total loop phase simply becomes \(\theta_{\mathrm{ser}}=\omega\tau+\varphi_{\mathrm{ps}}+\varphi_{\mathrm{YIG}}\). In this case, the phase shifter and YIG contributions enter as relatively smooth additive phase perturbations to the loop. Specifically, the phase perturbation enters linearly through \(\delta\theta_{\mathrm{ser}}=\delta\varphi_{\mathrm{ps}}+\delta\varphi_{\mathrm{YIG}}\). Consequently, the oscillator primarily experiences a smooth global phase shift with little internal interferometric amplification. The corresponding effective mismatch, 
\vspace{-8pt}
\begin{align}
\Delta_{\mathrm{eff}}=2\omega-\omega_p+\frac{1}{\tau_\mathrm{eff}} \delta\theta_{\mathrm{ser}}(\omega,\varphi_\mathrm{ps}), 
\end{align}
\vspace{-14pt}

\noindent therefore changes relatively gradually, allowing the OEPO to remain pinned over a broad perturbation range. Consequently, the inline structure lacks this interferometric enhancement mechanism because it contains only a single propagating path with no coherent branch competition or near-cancellation condition. The inline system behaves primarily as a globally phase-shifted oscillator, whereas the MZI behaves as a highly phase-sensitive interferometric eigenmode system.

\subsection{B. Analytial model for YIG-induced phase response}
\label{app:lorentz}

\renewcommand{\theequation}{B-\arabic{equation}}
\setcounter{equation}{0}  
\renewcommand{\thefigure}{B-\arabic{figure}}
\setcounter{figure}{0}  

\subsubsection{Complex Lorentzian model}

The field-dependent responses measured from the phase-shifter branch (CH1), the YIG branch (CH2), and their net difference (CH1$-$CH2), as shown in Fig.~\ref{Fig_YIG}, may be described using mixed symmetric--antisymmetric Lorentzian functions. For the phase-shifter branch, the response is written as: 
\vspace{-8pt}
\begin{align}
R_{\mathrm{ps}}(H)
&=
B_{\mathrm{ps}}
+
A_{\mathrm{ps}}
\Big[
\cos\psi_{\mathrm{ps}}\,
L_{\mathrm{sym}}(H)
\nonumber\\
&\qquad\qquad
+
\sin\psi_{\mathrm{ps}}\,
L_{\mathrm{asym}}(H)
\Big],
\end{align}
\vspace{-16pt}

\noindent while for the YIG branch: 
\vspace{-8pt}
\begin{align}
R_{\mathrm{YIG}}(H)
&=
B_{\mathrm{YIG}}
+
A_{\mathrm{YIG}}
\Big[
\cos\psi_{\mathrm{YIG}}\,
L_{\mathrm{sym}}(H)
\nonumber\\
&\qquad\qquad
+
\sin\psi_{\mathrm{YIG}}\,
L_{\mathrm{asym}}(H)
\Big].
\end{align}
\vspace{-16pt}

Here, \(B_{\mathrm{ps}}\) and \(B_{\mathrm{YIG}}\) denote the nonresonant backgrounds, \(A_{\mathrm{ps}}\) and \(A_{\mathrm{YIG}}\) are the fitted resonance amplitudes, and \(\psi_{\mathrm{ps}}\) and \(\psi_{\mathrm{YIG}}\) are the Lorentzian phases that determine the relative weights of the symmetric absorptive and antisymmetric dispersive components. The symmetric and antisymmetric basis functions are respectively given by
\vspace{-8pt}
\begin{align}
L_{\mathrm{sym}}(H)
&=
\frac{\Gamma^2}
{(H-H_0)^2+\Gamma^2},
\end{align}
\vspace{-16pt}

\noindent and
\vspace{-8pt}
\begin{align}
L_{\mathrm{asym}}(H)
&=
\frac{\Gamma(H-H_0)}
{(H-H_0)^2+\Gamma^2},
\end{align}
\vspace{-12pt}

\noindent where \(H_0\) is the resonance field and \(\Gamma\) is the linewidth parameter. In this framework, the symmetric Lorentzian component corresponds primarily to absorptive magnonic loss, whereas the antisymmetric component corresponds primarily to dispersive phase rotation. 

The differential response between the two branches may then be constructed as:
\vspace{-8pt}
\begin{align}
R_{\mathrm{diff}}(H)
=
R_{\mathrm{ps}}(H)
-
R_{\mathrm{YIG}}(H),
\end{align}
\vspace{-16pt}

\noindent which may again be fitted using a mixed Lorentzian form,
\vspace{-16pt}
\begin{align}
R_{\mathrm{diff}}(H)
&=
B_{\mathrm{diff}}
+
A_{\mathrm{diff}}
\Big[
\cos\psi_{\mathrm{diff}}\,
L_{\mathrm{sym}}(H)
\nonumber\\
&\qquad\qquad
+
\sin\psi_{\mathrm{diff}}\,
L_{\mathrm{asym}}(H)
\Big].
\end{align}
\vspace{-16pt}

Equivalently, assuming both branches share the same resonance field and linewidth, the differential response may be written explicitly as: 
\vspace{-8pt}
\begin{align}
R_{\mathrm{diff}}
&=
\Big(
A_{\mathrm{ps}}
\cos\psi_{\mathrm{ps}}
-
A_{\mathrm{YIG}}
\cos\psi_{\mathrm{YIG}}
\Big)
L_{\mathrm{sym}}
\nonumber\\
&\qquad
+
\Big(
A_{\mathrm{ps}}
\sin\psi_{\mathrm{ps}}
-
A_{\mathrm{YIG}}
\sin\psi_{\mathrm{YIG}}
\Big)
L_{\mathrm{asym}},
\end{align}
\vspace{-16pt}

\noindent such that the differential Lorentzian phase becomes: 
\vspace{-4pt}
\begin{align}
\tan\psi_{\mathrm{diff}}
&=
\frac{
A_{\mathrm{ps}}
\sin\psi_{\mathrm{ps}}
-
A_{\mathrm{YIG}}
\sin\psi_{\mathrm{YIG}}
}
{
A_{\mathrm{ps}}
\cos\psi_{\mathrm{ps}}
-
A_{\mathrm{YIG}}
\cos\psi_{\mathrm{YIG}}
}.
\end{align}
\vspace{-8pt}

\subsubsection{Amplitude--phase conversion in the adaptive oscillator}

The experimentally observed mixed symmetric--antisymmetric Lorentzian lineshapes measured near the splitter in Fig.~\ref{Fig_YIG}(f-h) indicate that the regenerated oscillation contains substantial amplitude--phase conversion generated through adaptive interferometric feedback redistribution. We construct a useful framework by treating the regenerated responses as linear projections of the local YIG absorptive and dispersive perturbations. The local YIG response may be decomposed into an absorptive amplitude perturbation \(\delta a_{\mathrm{YIG}}\) and a dispersive phase perturbation \(\delta\varphi_{\mathrm{YIG}}\), such that:
\vspace{-6pt}
\begin{align}
\delta a_{\mathrm{YIG}}
&\propto
L_{\mathrm{sym}}(H),
\\
\delta\varphi_{\mathrm{YIG}}
&\propto
L_{\mathrm{asym}}(H).
\end{align}
\vspace{-16pt}

The regenerated splitter-side amplitude and phase responses may then be written as: 
\vspace{-4pt}
\begin{align}
\begin{pmatrix}
\delta A_{\mathrm{splitter}} \\
\delta\varphi_{\mathrm{splitter}}
\end{pmatrix}
=
\begin{pmatrix}
C_{AA} & C_{A\varphi} \\
C_{\varphi A} & C_{\varphi\varphi}
\end{pmatrix}
\begin{pmatrix}
\delta a_{\mathrm{YIG}} \\
\delta\varphi_{\mathrm{YIG}}
\end{pmatrix},
\end{align}
\vspace{-8pt}

\noindent where \(C_{AA}\) describes direct amplitude preservation, \(C_{\varphi\varphi}\) describes direct phase preservation, \(C_{A\varphi}\) describes phase-to-amplitude conversion, and \(C_{\varphi A}\) describes amplitude-to-phase conversion generated through interferometric redistribution and feedback regeneration. Consequently, the regenerated splitter-side phase response becomes: 
\vspace{-6pt}
\begin{align}
\delta\varphi_{\mathrm{splitter}}
&=
C_{\varphi A}
L_{\mathrm{sym}}(H)
+
C_{\varphi\varphi}
L_{\mathrm{asym}}(H),
\end{align}
\vspace{-14pt}

\noindent which may be compared directly with the mixed Lorentzian fitting form:
\vspace{-8pt}
\begin{align}
R(H)
=
A
\Big[
\cos\psi\,
L_{\mathrm{sym}}(H)
+
\sin\psi\,
L_{\mathrm{asym}}(H)
\Big].
\end{align}
\vspace{-12pt}

The fitted Lorentzian phase therefore satisfies:
\vspace{-6pt}
\begin{align}
\tan\psi_{\varphi}
=
\frac{
C_{\varphi\varphi}
}{
C_{\varphi A}
},
\end{align}
\vspace{-12pt}

\noindent demonstrating that the measured Lorentzian phase directly quantifies the ratio between phase-preserving redistribution and amplitude-to-phase conversion inside the regenerated oscillator field. 

Physically, \(\psi_\varphi\approx0^\circ\) indicates that the measured phase arises primarily from amplitude-to-phase conversion, while \(\psi_\varphi\approx90^\circ\) indicates that the regenerated phase predominantly preserves the intrinsic local YIG dispersive response. Consequently, the experimentally observed mixed Lorentzian phases near the splitter directly quantify how local YIG absorption and dispersion become redistributed and converted into regenerated oscillator amplitude and phase through the adaptive MZI-OEPO loop.

Similarly, the regenerated amplitude response becomes: 
\vspace{-8pt}
\begin{align}
\delta A_{\mathrm{splitter}}
&=
C_{AA}
L_{\mathrm{sym}}(H)
+
C_{A\varphi}
L_{\mathrm{asym}}(H),
\end{align}
\vspace{-16pt}

such that the amplitude Lorentzian phase satisfies: 
\vspace{-8pt}
\begin{align}
\tan\psi_A
=
\frac{
C_{A\varphi}
}{
C_{AA}
}.
\end{align}
\vspace{-14pt}

Likewise, \(\psi_A\approx0^\circ\) indicates that the regenerated amplitude primarily reflects local absorptive microwave loss, whereas \(\psi_A\approx90^\circ\) indicates that the measured amplitude is dominated by phase-to-amplitude conversion through adaptive interferometric redistribution.

\subsection{C. YIG response at different PS bias}

\renewcommand{\theequation}{C-\arabic{equation}}
\setcounter{equation}{0}  
\renewcommand{\thefigure}{C-\arabic{figure}}
\setcounter{figure}{0}  

We also measured the phase evolution at other selective $\varphi_\mathrm{ps}$ values and for the respective sampling configurations, as summarized in Fig.~\ref{Fig_OEPO}. Consistent lineshapes were observed across a large $\varphi_\mathrm{ps}$ range. This is because the PS phase tunes how the YIG response is projected onto the global MZI-OEPO eigenstate, but it does not strongly modify the intrinsic local YIG susceptibility. As a result, the YIG-branch Lorentzian lineshape remains robust against PS bias, while the recombined output, differential signal, or global loop response may still show stronger PS-dependent modulation. The results support the interpretation that the YIG acts as a local resonant dispersive defect, while the PS acts as a global interferometric boundary-control knob.

\begin{figure*}[htb]
 \centering
 \includegraphics[width=6.6 in]{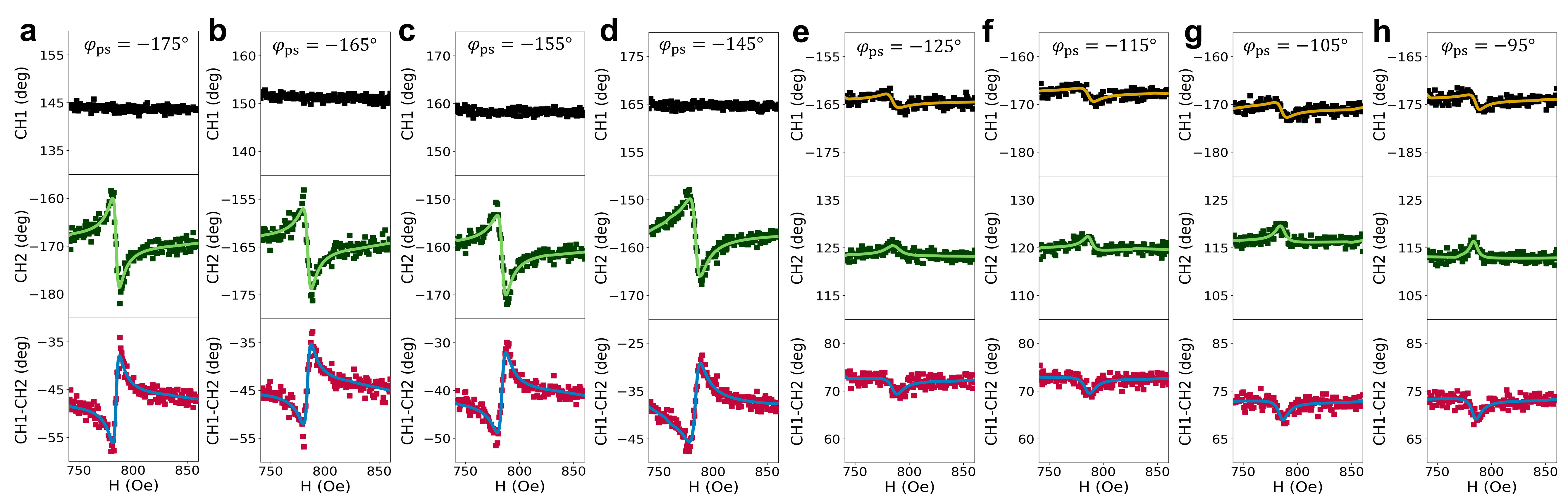}
 \caption{In an OEPO setup. (a-d) YIG-induced phase evolution of MZI branches sampled near the combiner at selective $\varphi_\mathrm{ps}$ values, (a) $-175^\circ$, (b) $-165^\circ$, (c) $-155^\circ$, (d) $-145^\circ$. (e-h) YIG-induced phase evolution of MZI branches sampled near the splitter at selective $\varphi_\mathrm{ps}$ values, (e) $-125^\circ$, (f) $-115^\circ$, (g) $-105^\circ$, (h) $-95^\circ$.}  
 \label{Fig_App_OEPO} 
\end{figure*}

A quantitative understanding of this YIG-branch independence may be obtained by distinguishing between local resonant variables and global interferometric eigenstate variables. Following earlier discussions around the Eq.~\ref{Eq_MZI}, the two branch transfer functions are respectively given by: 
\vspace{-8pt}
\begin{align}
H_{\mathrm{ps}}
&=
A_{\mathrm{ps}}
e^{i\varphi_{\mathrm{ps}}},
\\
H_{\mathrm{YIG}}
&=
A_{\mathrm{YIG}}(H)
e^{i\varphi_{\mathrm{YIG}}(H)}.
\end{align}
\vspace{-16pt}

The important distinction is that the magnetic-field sweep primarily modifies the local YIG susceptibility through \(A_{\mathrm{YIG}}(H)\) and \(\varphi_{\mathrm{YIG}}(H)\), whereas the phase-shifter sweep primarily modifies the global interferometric phase imbalance according to:
\vspace{-8pt}
\begin{align}
\Delta\varphi_{\mathrm{MZI}}
=
\varphi_{\mathrm{ps}}
-
\varphi_{\mathrm{YIG}}.
\end{align}
\vspace{-16pt}

Thus, the two perturbations belong fundamentally to different dynamical classes: the magnetic field controls a local resonant susceptibility perturbation, while the phase shifter controls a global interferometric boundary condition. In the OEPO configuration, the oscillation frequency remains strongly pinned at $f_p/2$, such that perturbations cannot be efficiently relaxed through frequency pulling or other spectral motion. Instead, the oscillator adaptively redistributes phase in order to satisfy the global phase-locking condition. 

Consequently, during a magnetic-field sweep, the dominant response is governed primarily by the intrinsic local YIG susceptibility \(\chi(H)\), which remains only weakly affected by the phase-shifter bias. Therefore, the experimentally observed coexistence of robust YIG Lorentzian lineshapes against phase-shifter bias naturally reflects the partial separation between local resonant susceptibility perturbations and globally constrained adaptive interferometric redistribution inside the phase-pinned MZI-OEPO architecture.

\subsection{D. Measurements in the OEO configuration}

\renewcommand{\theequation}{D-\arabic{equation}}
\setcounter{equation}{0}  
\renewcommand{\thefigure}{D-\arabic{figure}}
\setcounter{figure}{0}  

\begin{figure*}[htb]
 \centering
 \includegraphics[width=7 in]{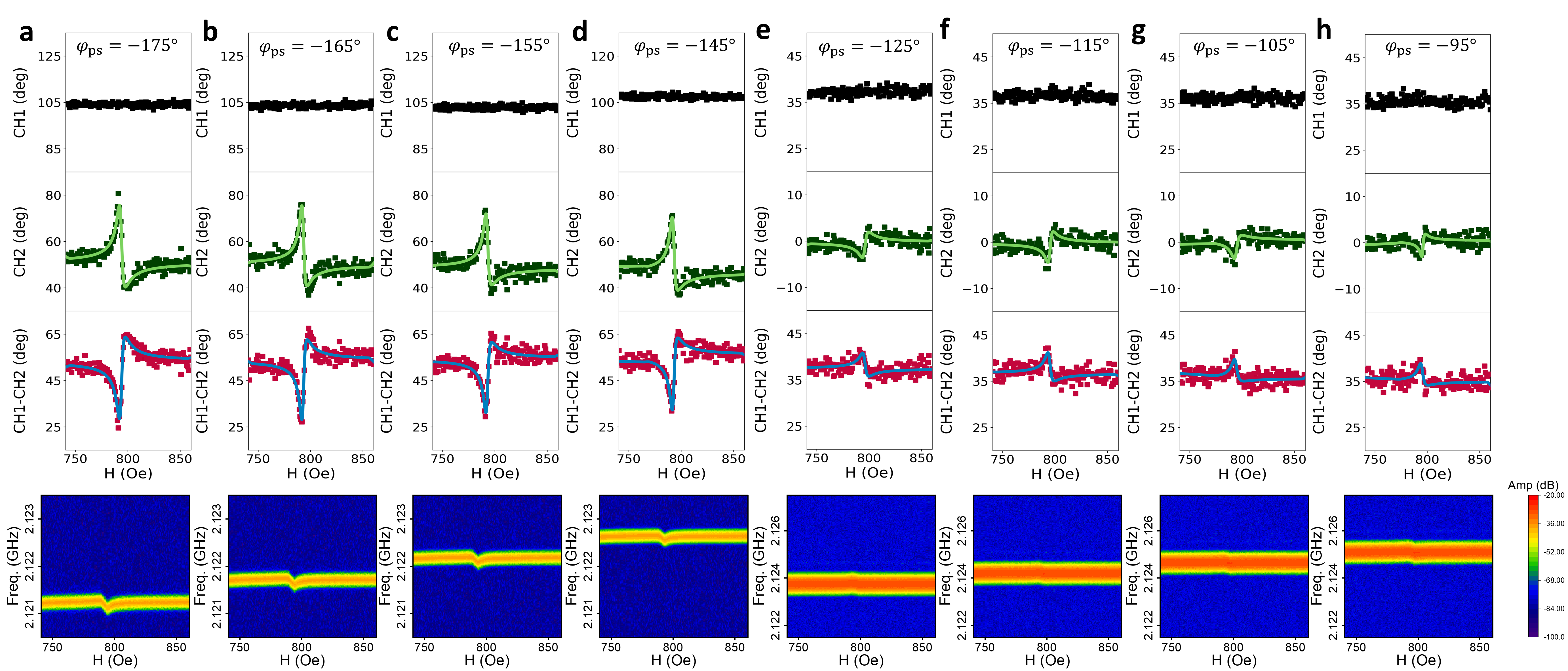}
 \caption{In an OEO setup. (a-d) YIG-induced phase evolution of MZI branches sampled near the combiner at selective $\varphi_\mathrm{ps}$ values, (a) $-175^\circ$, (b) $-165^\circ$, (c) $-155^\circ$, (d) $-145^\circ$. (e-h) YIG-induced phase evolution of MZI branches sampled near the splitter at selective $\varphi_\mathrm{ps}$ values, (e) $-125^\circ$, (f) $-115^\circ$, (g) $-105^\circ$, (h) $-95^\circ$. The last row show the corresponding spectra obtained by the spectrum analyzer, exhibiting nontrivial frequency-pulling due to the ability to self-adjust the eigenmode frequency in an OEO setup. }  
 \label{Fig_App_OEO}
\end{figure*}

We also measured the YIG-induced phase evolution for the MZI branches in a conventional OEO configuration by removing the external pump, see Fig.~\ref{Fig_App_OEO}. The same $\varphi_\mathrm{ps}$ values are selected as in the prior OEPO case (Fig.~\ref{Fig_App_OEPO}). Note that in this case, the frequency is no longer the pinned OEPO degenerate mode $f_p/2$, but instead, the generic OEO loop mode near 2.12 GHz, see Fig.~\ref{Fig_FSR}(b).    

In contrast to the OEPO configuration, where the locally sampled YIG response remains predominantly dispersive, the OEO configuration exhibits mixed symmetric--antisymmetric complex Lorentzian lineshapes at both the splitter-end and combiner-end sampling positions when sweeping the magnetic field $H$ across the YIG resonance, see the phase traces shown in Fig.~\ref{Fig_App_OEO}. In addition, the YIG-induced dispersive phase perturbation produces nontrivial adaptive frequency pulling in the OEO case, which can be directly monitored using the spectrum analyzer, see the bottom row of Fig.~\ref{Fig_App_OEO},  with an estimated variation $\sim1$ MHz across the FMR regime. 

Interestingly, both the phase trace and the frequency evolution (pulling) exhibits a mixed symmetric--antisymmetric complex Lorentzian lineshape, indicating that frequency pulling, adaptive gain redistribution, and amplitude--phase conversion become intertwined with the intrinsic YIG susceptibility response. Besides, the frequency pulling measured near the splitter end appears less pronounced, likely because the splitter-end signal probes the regenerated global oscillation after adaptive interferometric redistribution, where part of the local YIG-induced spectral perturbation has already been redistributed and partially compensated through coherent feedback and common-mode phase adaptation across the entire loop.

The experimentally observed distinction between the OEPO and OEO configurations may be understood from the fundamentally different ways in which the two oscillators respond to YIG-induced phase perturbations. In the OEPO configuration, the measured YIG phase response sampled after the YIG element remains predominantly dispersive because the oscillation frequency is strongly pinned by the parametric pump, at $f_p/2$, suppressing spectral relaxation and forcing the YIG-induced perturbation to remain primarily stored as local coherent phase evolution. Consequently, the measured phase response closely follows the intrinsic dispersive FMR susceptibility \(\delta\varphi_{\mathrm{YIG}}(H)\propto L_{\mathrm{asym}}(H)\). By contrast, in the herein OEO configuration the oscillation frequency is allowed to adapt through frequency pulling, such that the measured local phase becomes: 
\vspace{-8pt}
\begin{align}
\delta\varphi_{\mathrm{local}}(H)
\approx
\delta\varphi_{\mathrm{YIG}}(H)
+
2\pi\Delta\tau_{\mathrm{rel}}\delta f(H),
\end{align}
\vspace{-16pt}

\noindent where \(\Delta\tau_{\mathrm{rel}}\) is the relative branch delay imbalance and \(\delta f(H)\) is the YIG-induced OEO frequency pulling. Experimentally, \(\delta f(H)\) itself exhibits a mixed symmetric--antisymmetric complex Lorentzian character, implying that the pulling-induced correction simultaneously modifies both the absorptive and dispersive quadratures of the measured phase response. Consequently, the OEO pulling term can both renormalize the effective dispersive contribution and introduce additional symmetric Lorentzian admixture through adaptive spectral relaxation and interferometric redistribution. For a practical estimation, the phase correction induced by OEO frequency pulling may be estimated:
\vspace{-8pt}
\begin{align}
\delta\varphi_{\mathrm{pull}}
=
2\pi\tau_{\mathrm{rel}}\delta f. 
\end{align}
\vspace{-16pt}

For the experimentally observed frequency pulling of approximately \(\delta f\sim1~\mathrm{MHz}\), see the last row in Fig.~\ref{Fig_App_OEO}, the induced phase correction becomes directly proportional to the relative delay imbalance \(\tau_{\mathrm{rel}}\) between the two interferometric branches. Since the MZI branches were intentionally designed to remain approximately balanced, the experimentally relevant \(\tau_{\mathrm{rel}}\) is much smaller than the full loop delay extracted from the FSR.  For example, for small delays \(\tau_{\mathrm{rel}}\approx5-10~\mathrm{ns}\), the pulling-induced phase correction is approximately \(\delta\varphi_{\mathrm{pull}}\approx1.8-3.6^\circ\). Such perturbations are sufficiently large to substantially rotate the fitted Lorentzian phase \(\psi\) through redistribution between the symmetric and antisymmetric Lorentzian quadratures, as observed in Fig.~\ref{Fig_App_OEO}.

In this sense, the OEO behaves more like a self-adjusting frequency-tracking oscillator, whereas the OEPO behaves more like a phase-constrained coherent interferometric oscillator. Therefore, even when measured at the same downstream sampling location, the local YIG response naturally appears predominantly dispersive in the OEPO configuration but comparatively more absorptive in the OEO configuration.

\bibliography{ref}

\end{document}